\documentclass[%
 reprint,
superscriptaddress,
showpacs,
amsmath,amssymb,
aps,
prb,
floatfix,
]{revtex4-1}
\usepackage{graphicx}
\usepackage{subfigure} 
\usepackage[colorlinks,linkcolor=blue,anchorcolor=green,citecolor=blue]{hyperref}
\usepackage{bm}%

\begin{document}

\title{Enhanced electrocaloric cooling by electric field reversal}

\author{Yang-Bin Ma}

\email{y.ma@mfm.tu-darmstadt.de}
\affiliation{Institute of Materials Science, Technische Universit{\"a}t Darmstadt, 64287 Darmstadt, Germany}

\author{Nikola Novak}
\affiliation{Institute of Materials Science, Technische Universit{\"a}t Darmstadt, 64287 Darmstadt, Germany}

\author{Jurij Koruza}
\affiliation{Institute of Materials Science, Technische Universit{\"a}t Darmstadt, 64287 Darmstadt, Germany}

\author{Tongqing Yang}
\affiliation{College of Materials Science and Engineering, Tongji University, 200092 Shanghai, China}

\author{Karsten Albe}
\affiliation{Institute of Materials Science, Technische Universit{\"a}t Darmstadt, 64287 Darmstadt, Germany}

\author{Bai-Xiang Xu}
\affiliation{Institute of Materials Science, Technische Universit{\"a}t Darmstadt, 64287 Darmstadt, Germany}

\date{\today}
 
\begin{abstract}
An improved thermodynamic cycle is proposed, where the cooling effect of an electrocaloric refrigerant is enhanced by applying a reversed electric field. 
In contrast to conventional adiabatic heating or cooling by
on-off cycles of the external electric field, applying a reversed field is significantly improving the cooling efficiency, since the variation in configurational entropy is increased. By comparing results from computer simulations using Monte-Carlo algorithms and experiments using direct electrocaloric measurements, we show that the electrocaloric cooling efficiency can be enhanced by more than 20\% in standard ferroelectrics and also relaxor ferroelectrics, like Pb(Mg$_{1/3}$/Nb$_{2/3}$)$_{0.71}$Ti$_{0.29}$O$_3$.
\begin{description}
\item[PACS numbers]
77.70.+a, 77.80.-e, 77.80.Jk, 74.62.Dh 
\end{description}
\end{abstract}

\maketitle

%

Solid state refrigeration is a promising environment-friendly cooling technology based on the magnetocaloric, electrocaloric (EC) or elastocaloric effect~\cite{2012_Faehler}.
In the realm of EC refrigeration, much effort has been devoted to explore appropriate materials ~\cite{1999_Kutnjak,2014_Ozbolt,2014_Moya,2014_Correia,2006_Mischenko,2014_Pirc,2008_Neese,2015_Koruza,2013_Bai,2013_Moya,2009_Correia}
and device concepts,~\cite{2009_Epstein,2014_Gu,2014_Alpay} while
theoretical studies helped to understand the physical mechanism of the electrocaloric effect (ECE).~\cite{2009_Cao,2011_Dunne,2011_Pirc,2012_Rose}
There is, however, much less work on optimal EC cycles.~\cite{2012_Ponomareva}

In a conventional EC cycle of a solid refrigerant, the cooling effect is obtained simply by removing the previously applied electric field.~\cite{2011_Scott} Ferroelectric materials typically exhibit a positive ECE:
by applying an external field the temperature increases, whereas removal of the field decreases the temperature. 
Antiferroelectric materials can show a negative ECE and the temperature response is inverse,~\cite{2014_Pirc} because configurational entropy is growing if an external field is applied, which is causing a temperature drop. Similarly, a negative ECE has also been observed in relaxor ferroelectrics~\cite{2010_Perantie}, and based on computer simulations Ponomareva and Lisenkov~\cite{2012_Ponomareva} recently showed that a transition between a positive and negative ECE is possible if polarization and electric field become non-collinear. 
Experimental and theoretical studies of adiabatic loading cycles in PZT~\cite{1968_Thacher,2013_Wang,2014_Zeng} showed occurrence of a negative EC under reversed electric field. This implies, that field reversal could enhance the ECE.

In this paper, we show that in an ECE cycle the EC cooling effect is enhanced when transferring from the isothermal to adiabatic stage, if the previously applied electric field is not only removed but reversed. This is because single-phase ferroelectric materials, having a distinct hysteresis effect,
exhibit a negative ECE in prepoled samples with remnant polarization. 
Thus, the temperature can further drop, thanks to this negative ECE. 
By comparing results from computer simulations and experiment, we demonstrate that there is an optimal magnitude of the reversed field in order to maximize the ECE.

\begin{figure}[!htbp]
  \centering
  \centerline{\includegraphics[width=7cm]{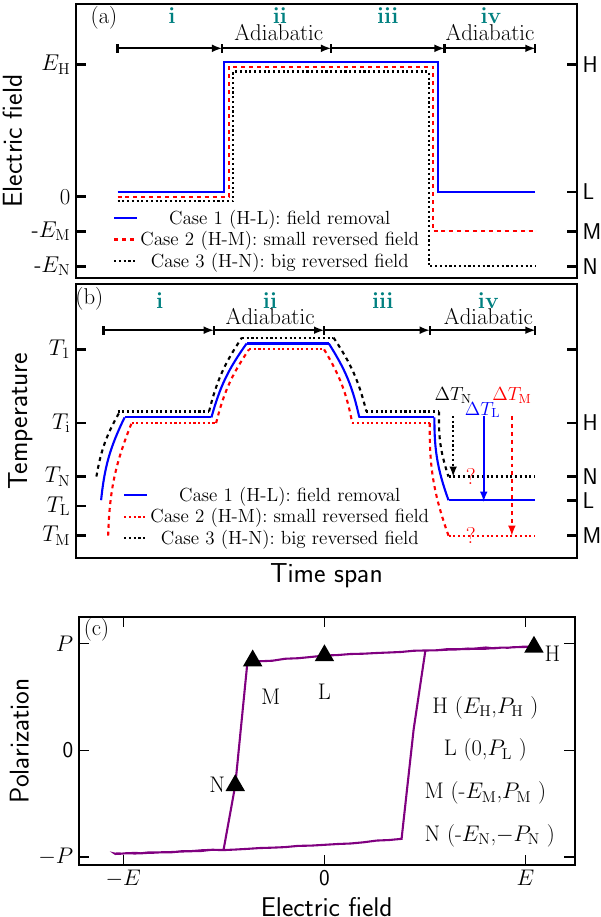}}
   \caption{
   The loading history (a), the possible temperature change (b) and the corresponding polarization change (c) are sketched for three different cases. 
The cooling procedure, i.e., the temperature drop from step $\mathbf{\romannumeral 3}$ to step $\mathbf{\romannumeral 4}$, is of interest to the current work.
For case 1, at step $\mathbf{\romannumeral 4}$ after removing the field the macroscopic polarization decreases (H to L in (c)). 
Hence, the electrocaloric cooling is achieved with the temperature drop $\Delta T_\mathrm{L}$ (see (b)).
For case 2, a small reversed electric field is applied (H to M), while for case 3 a high field is applied (H to N).
The resultant temperature drop might be enhanced ($\Delta T_\mathrm{M}$) or degenerated ($\Delta T_\mathrm{N}$), which is judged by the simulation and experiment. 
   }
 \label{fig:figure1}
\end{figure}


An EC cycle contains the following four steps: isothermal heat absorption, adiabatic heating, isothermal heat release, and adiabatic cooling. 
In following we compare the temperature changes obtained for three distinct scenarios, which are depicted in Fig.~\ref{fig:figure1}(a,b). Contrary to existing studies,~\cite{1968_Thacher,2013_Wang,2014_Zeng} here we are probing a sequence of isothermal and adiabatic steps as they would occur in a cooling device. 
The first case corresponds to a conventional EC cycle, where the cooling effect is obtained by removing the applied field. In the P-E loop, this corresponds to the change of polarization from the poled state H to the remnant state L shown in Fig.~\ref{fig:figure1}(c). In the other two cases, we study the influence of a small and large reversed field, respectively. As it is illustrated in Fig.~\ref{fig:figure1}(c), the small reversed field lies in the range betwen point L and M, where no significant polarization switching takes place, while the large reversed field $E_N$ corresponds to the state N, where macroscopic polarization switching has started.

In our simulations, we use lattice-based Monte-Carlo simulations in the canonical and microcanonical ensemble based on a Ginzburg-Landau type Hamiltonian. Details of the model parametrized for tetragonal BaTiO$_3$ and the simulation setup are described in Ref.~\onlinecite{2015_Ma,2015_Maa}. 
The initial temperature is $T_0= 320$K in all simulations and samples were poled by a field of 65.8\,kV\,mm$^{-1}$. Thus, all cases start with an identical fully-poled configuration. 

\begin{figure}[htbp]
  \centering 
  \centerline{\includegraphics[width=7cm]{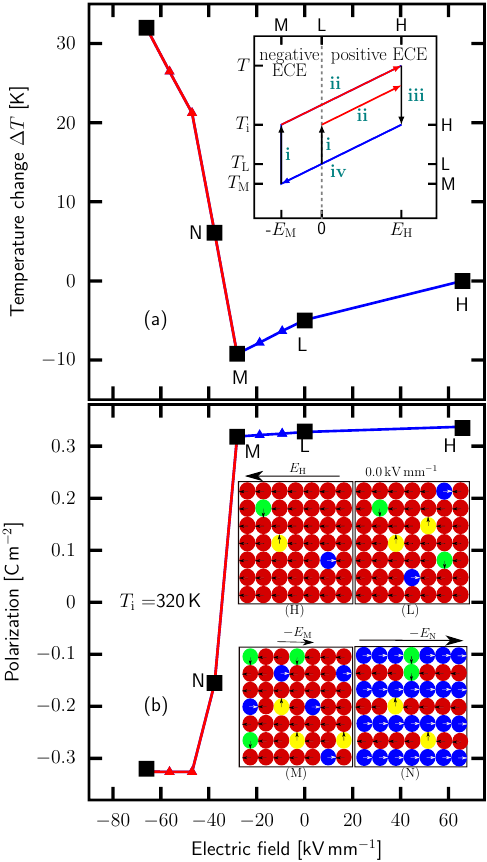}}
   \caption{
The prepoled field $E_H$=65.8\,kV\,mm$^{-1}$ is kept constant for all cases, and the same polarization configuration can be assured.
The polarization (a) and the temperature change (b) are depicted with respect to different reversed fields.
As shown in (a) and (b), with increasing the magnitude of the reversed field, the electrocaloric cooling is initially enhanced, but later degenerated.
The phenomena at points H, L, M and N in (a) and (b) are explained by the schematic domain patterns in the inset of (a).
At point H, the external field directs to the left, while at points M and N the reversed field directs to the right.
From H to L and M the polarization disorder decreases, and the configurational entropy $S_{conf}$ increases. 
This reveals that the application of the reversed field leads to the further decrease of the temperature.
However, from M to N, there is a sharp change of the polarization, and thus it is an irreversible adiabatic process. 
This leads to a huge increase of the total entropy and simultaneously the temperature.
The cooling and heating phases are mapped separately in blue and red colors.
The inset in (b) shows that compared with the conventional cycle (the interior cycle) the same negative electrocaloric effect (the exterior cycle) can be utilized to enhance both the cooling and heating effect.
Symbols $\mathbf{\romannumeral 1}$, $\mathbf{\romannumeral 2}$, $\mathbf{\romannumeral 3}$ and $\mathbf{\romannumeral 4}$ represent the same steps as described in Fig.~\ref{fig:figure1}.
}
   \label{fig:figure2}
\end{figure}

Results of the polarization and temperature changes are presented in Fig.\ref{fig:figure2}. 
As shown in Fig.~\ref{fig:figure2}(a), a temperature drop by $\Delta$ T= -5.0\,K is observed for the case of field removal (H $\rightarrow$ L).
An additional temperature drop leading to $\Delta T$= -8.9\,K can be achieved, if a small reversed field $|E_M| \leq 28.2$\,kV\,mm$^{-1}$ is applied (H $\rightarrow$ M). This indicates an enhanced EC cooling due to this reserved field, which is attributed to the negative ECE.

On the other hand, if a large reversed field of $|E_N| \leq 37.6$\,kV\,mm$^{-1}$ is applied, there is no temperature drop. Instead, a temperature increase relative to the starting temperature by $\Delta T$=6.1\,K is observed (H 
$\rightarrow$ N).

The schematic domain patterns for the initially poled state H, the remnant state L, the state M under small reversed field, and the switching state N are shown in inset of Fig.~\ref{fig:figure2}(a).
An increase of polarization disorder can be detected from H to L, as well as from L to M. 
Correspondingly, the polarization decreases slightly in both processes (see Fig.\ref{fig:figure2}(b)). 
These two processes are predominately reversible.
Thus, the total entropy $S_{total}=S_{conf} + S_{vib}$, including the configurational entropy $S_{conf}$ and the vibrational entropy $S_{vib}$, stays constant. 
Due to the elevated polarization disorder, $S_{conf}$ increases continuously from H to L and to M. It can be hence inferred that $S_{vib}$ deceases gradually. Thus, $T$ decreases from H to L, and further from L to M.~(see Fig.~\ref{fig:figure2}(a)).

If we increase the field and move from $H$ to $N$, the polarization switches significantly under the influence of the strong reversed field (see also Fig.\ref{fig:figure2}(b)).
This indicates an irreversible adiabatic process, which leads to a giant upsurge of total entropy, and an increase of both $S_{vib}$ and $S_{conf}$.
Therefore, a deterioration of the ECE is expected (see Fig.~\ref{fig:figure2}(a)).
Around the coercive field $S_{conf}$ is maximal, and the macroscopic polarization is zero. However, due to the irreversibility-induced increase of the total entropy, $S_{vib}$ does not take its minimum. In fact, from $S_{vib} = S_{total} - S_{conf}$, one concludes that the minimum of $S_{vib}$, i.e., the maximal temperature drop, occurs at the reversed electric field, which satisfies
\[
\dfrac{dS_{vib}}{dE} = \dfrac{dS_{total}}{dE} - \dfrac{dS_{conf}}{dE} = 0.
\]
This optimal reversed electric field should be smaller than the coercive field. This conclusion contrasts with the old report~\cite{1968_Thacher}, in which the Maxwell relation that is only valid for the reversible process was utilized for a process with considerable irreversibility.

It is noted that the same negative ECE can be also utilized to enhance the heating effect. More specifically, by removing the reversed electric field, the temperature can be increased, as it is illustrated in the inset of Fig.~~\ref{fig:figure2}(b).

\begin{figure}[htbp]
  \centering 
  \centerline{\includegraphics[width=7.9cm]{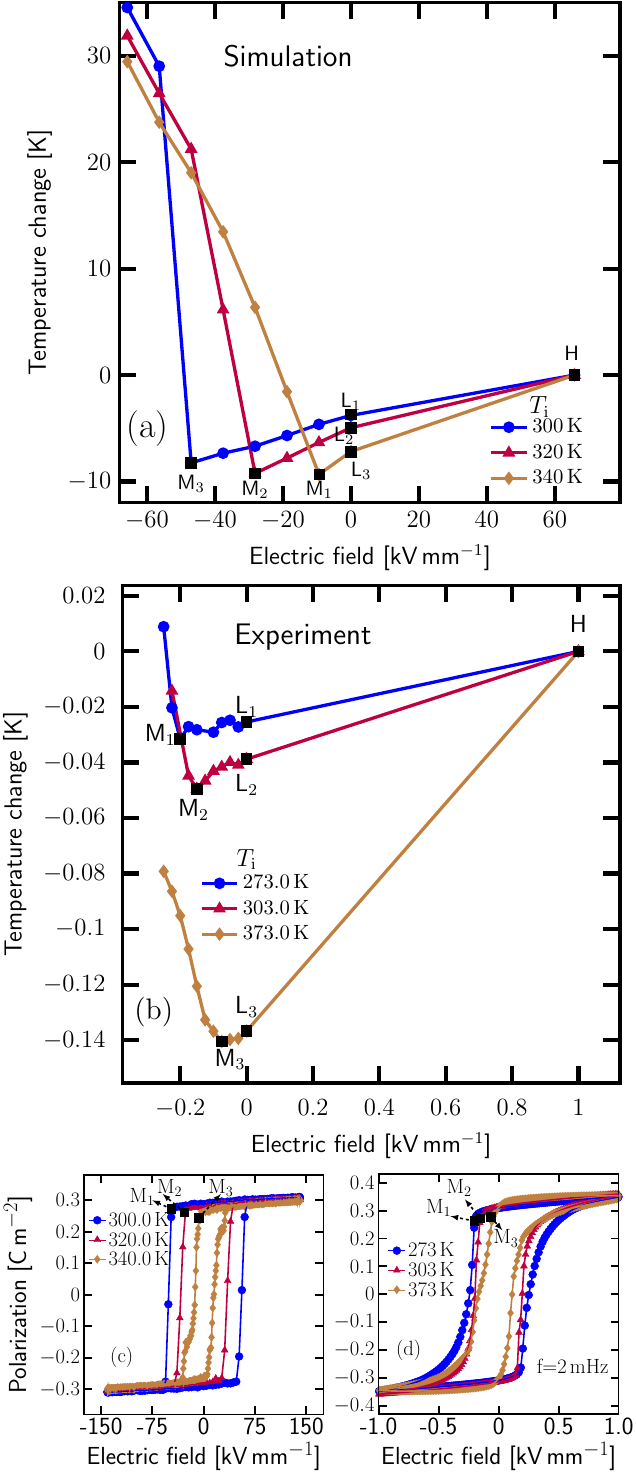}}
   \caption{
   Firstly, both the simulation (a) and the experiment (b) assert that with increasing the magnitude of the reversed field, the electrocaloric cooling is initially enhanced, and later deteriorated.
Secondly, with increasing the initial temperature $T_\mathrm{i}$, the coercive field decreases.
Therefore, the optimal reversed field with maximum temperature drop decreases.
The position of the maximum cooling effect is labeled with M$_1$, M$_2$ and M$_3$. 
The corresponding hysteresis loops (c) for the simulation and (d) for the experiments are shown.
The 2D idealized defect-free BaTiO$_3$ and the single crystal of Pb(Mg$_{1/3}$/Nb$_{2/3}$)$_{0.71}$Ti$_{0.29}$O$_3$ (PMN-29PT) are individually investigated in the simulation and the experiment, which leads to the quantitative differences.
}  
  \label{fig:figure3}
\end{figure}

In order to understand the influence of the initial temperatures $T_\mathrm{i}$ on the phenomenon of EC enhancement, results for three initial temperatures, $T_0 $ = 300\,K, 320\,K and 340\,K, are also compared. (see Fig.~\ref{fig:figure3}(a)) Note that all three initial temperatures are far below the Curie temperature 393\,K, where only the tetragonal phase is present.
At lower temperature, the coercive field is higher as confirmed by the simulated P-E loops shown in the inset of Fig.~\ref{fig:figure3}(a). 
Therefore, by increasing the initial temperature from 300\,K to 320\,K and 340\,K, the optimal reserved field with maximum EC cooling decreases.



It should be noted that the field reversal is only enhancing the ECE if the material is in its ferroelectric state, since in the paraelectric configuration irreversible contributions are missing.


For the experimental verification of the predicted EC response 
under the influence of reversed electric fields, direct EC measurement were carried out on Pb(Mg$_{1/3}$/Nb$_{2/3}$)$_{0.71}$Ti$_{0.29}$O$_3$ (PMN-29PT) single crystals. 
A platelet-shaped PMN-29PT single crystal was cut perpendicular to the [001] direction and polished. 
The geometry of platelet was 3.69$\times$4.35$\times$0.8\, mm$^3$. 
Surfaces were covered by sputtered silver electrodes on which cooper contact wires were attached with electrically conducting silver paste. 
In addition, a thermistor was attached on one side of the sample with electric non-conducting varnish to perform direct EC measurement. 
Prior to measurement the sample was heated up to 450\,K for 10\,min and then stabilized within 1\,mK at the target temperature. 
In the polarization hysteresis-loop measurements, the electric field is slowly cycled linearly with the frequency of 0.002\,Hz between $\pm$1\,kV\,mm$^{-1}$. 
The corresponding polarization charge was measured by a Keithley 6517B electrometer. 
The EC temperature change in PMN-29PT [001] was measured via a direct EC method.~\cite{2010_Rozic, 2013_Novaka} 
Each EC measurement consists of two parts, i.e., a heating part when the electric field is applied and a cooling part when the electric field is removed instantaneously. 
The actual EC temperature change $\Delta T$ was determined from the cooling part. 
The phenomena in the experiments agree with those revealed in the simulation qualitatively.
As can be seen in Fig.~\ref{fig:figure3}(b), with increasing the initial temperature from 273\,K to 303\,K and 373\,K, the optimal reversed field with maximum EC cooling decreases, as observed in the simulation.
At a certain $T_\mathrm{i}$ the EC cooling is firstly enhanced by increasing the reversed field from point L to M and subsequently reduced, if the reversed field is further increased. It should be noted that there is no phase transition in the temperature regime studied here, since PMN-29PT does only exist in the rhombohedral phase between 273 and 373\, K.

In summary, the experimental results confirm the predictions of the simulations in two aspects. Firstly, EC cooling is enhanced by applying a reserved field, and secondly, the optimal reserved electric field strength decreases with increasing initial temperature. 
It should be noted that a quasistatic loading scenario is considered in both, simulations and measurements. This assumption is justified, since the characteristic polarization switching time is on the order of 10$^{-6}$\,s~\cite{1954_Merz} and thus much faster than the typical EC measurement time scale of seconds.~\cite{2010_Kar-Narayan,2013_Moya}. In the Monte-Carlo simulations the magnitude of the applied electric field and the obtained temperature variation are higher than those obtained by the experimental results. This quantitative discrepancy is related to the fact that the simulation is performed with a 2 dimensional model of single crystalline defect-free BaTiO$_3$, while the experiments are done for the bulk single crystal of PMN-29PT. However, the underlying physics is maintained, i.e., the ECE can be enhanced by a proper magnitude of reversed field, thanks to the negative ECE.

We conclude that in ferroelectric materials, exhibiting a broad square-like isothermal hysteresis loop, the EC cooling can be greatly enhanced by field reversal. 
In contrast, irreversible adiabatic processes dominate in EC cycles in ferroelectric materials with slim hysteresis and thus an enhancement of the ECE can be hardly observed.
Therefore, the concept of applying a proper reversed field to enhance the EC cooling might be especially beneficial for thin-films, which have a square-shaped P-E loop.

The funding of Deutsche Forschungsgemeinschaft (DFG) SPP1599 (XU 121/1-2, AL 578/16-2, NO 1221/2-1) is gratefully acknowledged.
Competence Center of HPC Hessen is appreciated for computation time.
%

%

\bibliographystyle{apsrev4-1.bst} 

\end{document}